\RequirePackage{fix-cm}
\documentclass[smallcondensed]{svjour3}     
\smartqed  
\usepackage{graphicx}
\usepackage[numbers,square,sort]{natbib}
\begin{document}

\title{Non-pooling Network for medical image segmentation
}


\author{Weihu Song          \and
        Heng Yu              \and
        Jianhua Wu
}


\institute{Beihang University\at
              \email{weihusong@buaa.edu.cn}   \at
              Beijing, 100191, China
          \and
          Carnegie Mellon University \at
            \and
            Nankai University\at
}

\date{Received: date / Accepted: date}

\maketitle

\begin{abstract}
Existing studies tend to focus on model modifications and integration with higher accuracy, which improve performance but also carry huge computational costs, resulting in longer detection times. In medical imaging, the use of time is extremely sensitive. And at present most of the semantic segmentation models have encoder-decoder structure or double branch structure. Their several times of the pooling use with high-level semantic information extraction operation cause information loss although there is a reverse pooling or other similar action to restore information loss of pooling operation. In addition, we notice that visual attention mechanism has superior performance on a variety of tasks. Given this, this paper proposes non-pooling network(NPNet), non-pooling commendably reduces the loss of information and attention enhancement module effectively increases the weight of useful information. The method greatly reduces the number of parameters and computation costs by the shallow neural network structure. We evaluate the semantic segmentation model of our NPNet on three benchmark datasets comparing with multiple current state-of-the-art(SOTA) models, and the implementation results show that our NPNet achieves SOTA performance, with an excellent balance between accuracy and speed.
\end{abstract}

\section{Introduction}\label{sec1}
Medical image segmentation can promote the research and development of medical field. It can help doctors analyse and take action using image features. The accuracy and speed of image segmentation is critical and existing research is carried out from these two aspects. However, the relationship between accuracy and speed in most models has not reached a relative balance.
FCN\cite{Shelhamer2017FullyCN}, as the first semantic segmentation model, undoubtedly attracted great attention. It changed the last full connection layer of the classification network into convolution to achieve remarkable performance on semantic segmentation. U-net\cite{Ronneberger2015UNetCN}, the first segmentation network proposed for medical images, is a typical encoder-decoder structure. In U-net, skip connections are used to effectively integrate shallow spatial information and deep semantic information, thus making up for the loss of feature information caused by multiple pooling operations in the encoder stage. Subsequently, a series of improved models based on U-net show up. More complex feature extraction modules are used to extract as much feature information as possible from each level of the segmentation network to weaken the influence of pooling operations on information loss. Although such models improve certain performance, However, more redundant information, even error information, was introduced, and the model size and computation cost also increased significantly, creating a certain burden. SegNet\cite{Badrinarayanan2017SegNetAD} uses two classification networks as encoder and decoder respectively and proposes to use max pool index to do up-sampling to better restore the impact of pooling. PSPNet\cite{Zhao2017PyramidSP} and DeePLab series\cite{Chen2015SemanticIS}\cite{Chen2017RethinkingAC}\cite{Chen2018DeepLabSI}\cite{Chen2018EncoderDecoderWA} both use image classification networks as the backbone. The former proves the effectiveness of extracting multi-size feature information by pyramid pooling module for the first time, while the latter proposes to use dilated convolution to obtain feature information of larger receptive field, using dilated spatial pyramid pooling (ASPP) to obtain rich feature information of multiple dimensions. In addition, the attention mechanism introduced from NLP to computer vision has also shown its dazzling brilliance, among which SENet\cite{2017Squeeze} is undoubtedly the most important representative, its excellent performance won the last imagenet champion. Some other semantically segmented networks use dual branching structures\cite{han2020contextnet} to acquire semantic information, spatial information, and cascade structures\cite{2020CascadePSP} respectively.
Above all, most of the semantic segmentation models have encoder-decoder structure, use image classification network as the backbone,  and use the structure of the double branch or cascade structure. Pooling operation used in these semantic segmentation models leads to information loss. And the complex structure will also cause the burden of model and calculation. In addition, dilated convolution, multi-dimensional feature extraction, and attention mechanisms are proved to be effective. Therefore, we elaborately designed a simple and novelty non-pooling network, which solves the information loss caused by the pooling operation and uses improved ASPP and a new plug-and-play attention mechanism module.
Our contribution is in four aspects: (1) We propose a new plug-and-play attention mechanism module, which has better performance than the attention module in SENet. (2) We propose an improved ASPP module and have better performance. (3) For the first time, non-pooling semantic segmentation model with only 1/50 of the number of model parameters and computation costs of U-net is proposed. (4) Our method surpasses other state-of-the-art performance on three medical image segmentation datasets.

\section{Methodology}\label{sec3}
\begin{figure}[htp]

\begin{minipage}[b]{1.0\linewidth}
  \centering
  \centerline{\includegraphics[width=8.5cm]{ 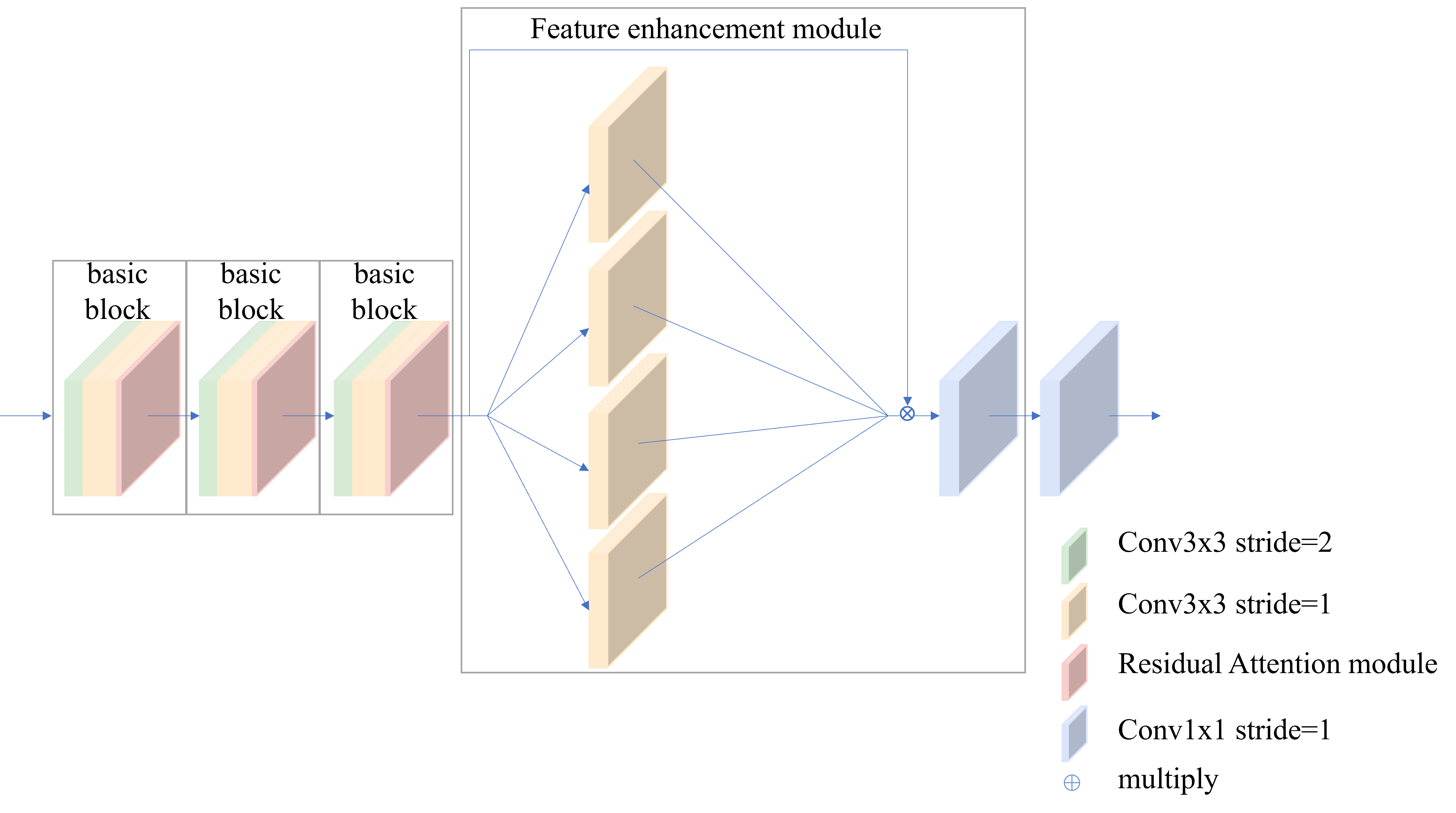}}
  \label{fig.NPNet}
  \centerline{(a) NPNet}\medskip
  
\end{minipage}
\begin{minipage}[b]{1.0\linewidth}
  \centering
  \centerline{\includegraphics[width=8.5cm]{ 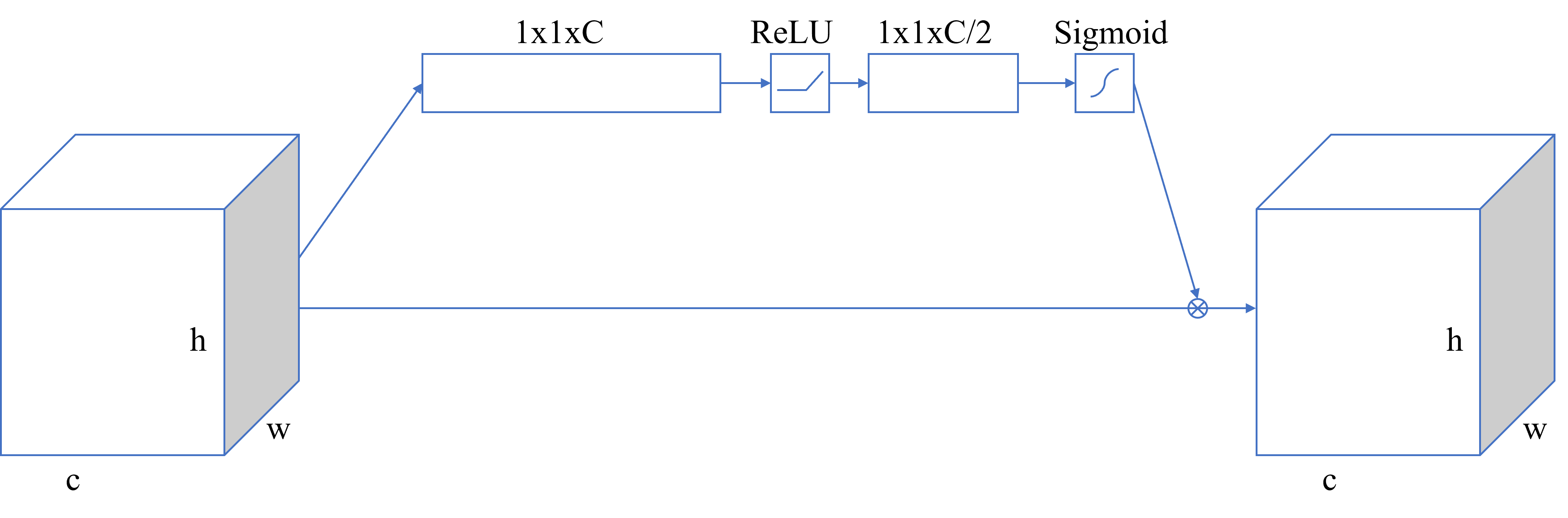}}
  \label{fig.att}
  \centerline{(b) attention enhancement module}\medskip
\end{minipage}

\caption{Proposed NPNet architecture.}
\label{network}
\end{figure}

In this paper, we propose NPNet, a novel lightweight semantic segmentation model for medical images. The network structure is described in Fig~\ref{network}. It is mainly composed of the basic block, attention enhancement module shown in Fig~\ref{network}, and feature enhancement module. In this model, all the convolution operations are 3x3 convolution kernel, followed by batch normalization(bn) and ReLU. There are three basic blocks at the beginning of the network, and attention enhancement module is added after each block, and a feature enhancement module is implemented in the middle of the network. At the back of the network, 1x1 convolution is used to output according to the classification number and bi-linear interpolation is used to restore the original input size. In this section, we will talk about these components in detail.

\subsection{Basic block}
The basic block first uses 3x3 convolution operation with stride equal to 2, which is equivalent to the function of reducing image size and computation achieved by the pooling operation of stride equal to 2. Moreover, the convolution operation can also effectively deal with the loss of feature information in pooling operation. Then, two 3x3 convolution operations with stride equal to 1 are used to fully extract the feature information of this layer, and also to obtain more abundant and useful feature information which is conducive to the transmission of the information of the next layer. The design of this basic block is to reduce the information loss and realize the effective extraction of spatial feature information of images with different sizes.

\subsection{Attention enhancement module}
Attention enhancement module is an attention module integrated with 1x1 convolution. First, the input image is transformed into a 1-dimensional matrix by adaptive average pooling, and then the input dimension is transformed into the input dimension divided by parameter reduction using 1x1 convolution. Like SENet, we also use linear activation function ReLU and nonlinear activation function Sigmoid successively. The difference is that we use 1X1 convolution to replace the full connection layer. 1X1 convolution can effectively improve the nonlinear characteristics and information interaction across channels, better extracting useful information in feature information. Finally, the nonlinear activation function Sigmoid is used on output and the result is weighted by multiplying the original input to achieve channel adaptive weighting. This module is used after each basic block to further strengthen the weight of useful information between different channels, thus providing better characteristic information for the following steps. 

\subsection{Feature enhancement module}
The feature enhancement module is used after three basic blocks, and the image size currently is 1/8 of the input size. This module is composed of four dilated convolutions with the ratio of 1, 5, 15, 20, and two 1x1 convolutions. The dilated convolution can obtain feature information from the large receptive field without increasing the number of parameters, obtaining richer semantic information. The input image is first output through four dilated convolutions according to 1/2 of the number of output channels. Then the four outputs are superimposed through concatenation, and the number of channels is 2 times the number of output channels. 1x1 convolution is used to achieve dimension reduction, that is, the normal number of output channels is obtained. At this point, the residual structure is introduced to concatenate the result with the original input information to realize feature reuse. Finally, 1x1 convolution is used for further dimension reduction.

\section{Experiments and Results}
\subsection{Datasets}

\textbf{Lung Segmentation.} Lung CT image segmentation is an important and initial step in lung CT image analysis. This dataset comes from the Kaggle contest, Finding and Measuring Lungs in CT data\footnote{https://www.kaggle.com/kmader/finding-lungs-in-ct-data}(luna for short). It consists of 267 2D images and is randomly split into train set (80\%) and test set (20\%). Also, we use the original image equally.\\
\textbf{Skin Lesion Segmentation.} Computer-aided automatic diagnosis of skin cancer is an inevitable trend, and skin lesions segmentation as the first step is urgent. The data set is from MICCAI 2018 Workshop - ISIC2018: Skin Lesion Analysis Towards Melanoma Detection\cite{codella2019skin}\cite{tschandl2018ham10000}(skin for short). It contains 2594 images and is randomly split into train set (80\%) and test set (20\%). For better model training and result display, we resize all the original images to 224 × 224.\\
\textbf{Polyp Segmentation.} Accurate detection of colon polyps is of great significance for the prevention of colon cancer. CVC-ClinicDB\cite{Bernal2015WMDOVAMF}(CVC for short) includes 612 colon polyp images. We use the original size 384x288 of image and split it into train set(80\%) and test set(20\%).

\subsection{Experimental Settings}
For three benchmarks and multiple segmentation models, we set consistent training parameters. we set epochs as 100 in the three data sets. We use a learning rate(LR) equal to 1e-3 for Luna and Skin task and 1e-4 for CVC task. In addition, we use batch size equal to 2 for luna and CVC task, and 4 for the skin task. Cross entropy loss and Adam are used as loss function and optimizer, respectively. All experiments run on the NVIDIA GeForce RTX 2080Ti GPU with 12GB. Intersection over Union (IOU), dice coefficient and computational complexity(MACs) are selected as the evaluation metrics in this paper. We used these evaluation metrics for all datasets.

\begin{table*}[htp]
    \caption{Comparsion on CVC, skin and luna with seven models}
    \centering
    \begin{tabular}{cccccccc}
        \hline
        Dataset& Methods& IOU& Dice& Params(M)& MACs(G)\\
        \hline
        CVC  & FCN8s\cite{Shelhamer2017FullyCN} & 0.6149 & 0.7249 & 14.72 & 33.89\\
        & SegNet\cite{Badrinarayanan2017SegNetAD} & 0.7146 & 0.7933 & 29.44 & 67.67\\
        & PSPNet\cite{Zhao2017PyramidSP} & 0.7159 & 0.8045 & 17.5 & 133.23\\
        & U-Net\cite{Ronneberger2015UNetCN} & 0.7439 & 0.8229 & 34.53 & 110.46\\
        & Attention U-Net\cite{Oktay2018AttentionUL} & 0.7334 & 0.8153 & 34.87 & 112.27\\
        & U-Net++\cite{Zhou2018UNetAN} & 0.7632 & 0.8356 & 36.63 & 233.88\\
        & NPNet & \textbf{0.7766} & \textbf{0.8397} & \textbf{0.71} & \textbf{2.17}\\
        \hline
        skin  & FCN8s & 0.7828 & 0.8511 & 14.72 & 61.50\\
        & SegNet & 0.7897 & 0.8558 & 29.44 & 30.71\\
        & PSPNet & 0.8052 & 0.8708 & 17.5 & 60.45\\
        & U-Net & 0.8086 & 0.8691 & 34.53 & 50.12\\
        & Attention U-Net & 0.8028 & 0.8691 & 34.87 & 50.94\\
        & U-Net++ & 0.7901 & 0.8588 & 36.63 & 106.11\\
        & NPNet & \textbf{0.8170} & \textbf{0.8757} & \textbf{0.71} & \textbf{0.99}\\
        \hline
        luna  & FCN8s & 0.9741 & 0.9802 & 14.72 & 80.32\\
        & SegNet & 0.9688 & 0.9789 & 39.76 & 160.41\\
        & PSPNet & 0.9732 & 0.9823 & 34.88 & 315.87\\
        & U-Net & 0.9749 & 0.9821 & 39.09 & 261.64\\
        & Attention U-Net & 0.9698 & 0.9794 & 29.30 & 266.11\\
        & U-Net++ & 0.9746 & 0.9831 & 36.23 & 554.37\\
        & NPNet & \textbf{0.9785} & \textbf{0.9832} & \textbf{0.71} & \textbf{5.15}\\
        \hline       
    \end{tabular}
    \label{table2}
\end{table*}

\subsection{Experimental Results}
\begin{figure*}[htp]

\begin{minipage}[b]{1.0\linewidth}
  \centering
  \centerline{\includegraphics[width=12cm]{ 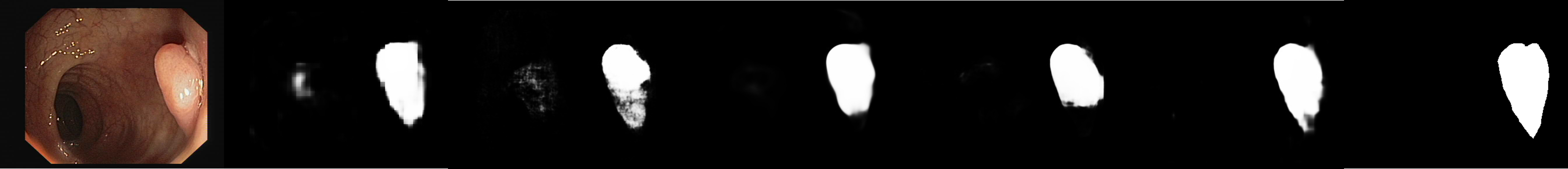}}
  \centerline{(a) CVC}\medskip
\end{minipage}
\begin{minipage}[b]{1.0\linewidth}
  \centering
  \centerline{\includegraphics[width=12cm]{ 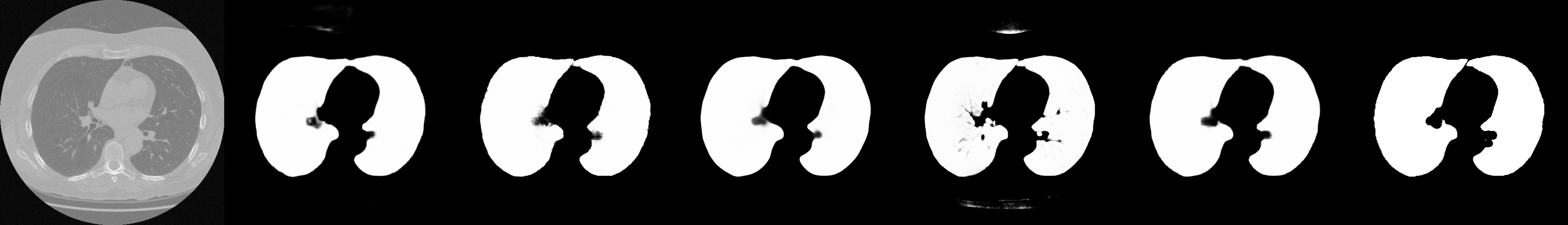}}
  \centerline{(b) luna}\medskip
\end{minipage}

\begin{minipage}[b]{1.0\linewidth}
  \centering
  \centerline{\includegraphics[width=12cm]{ 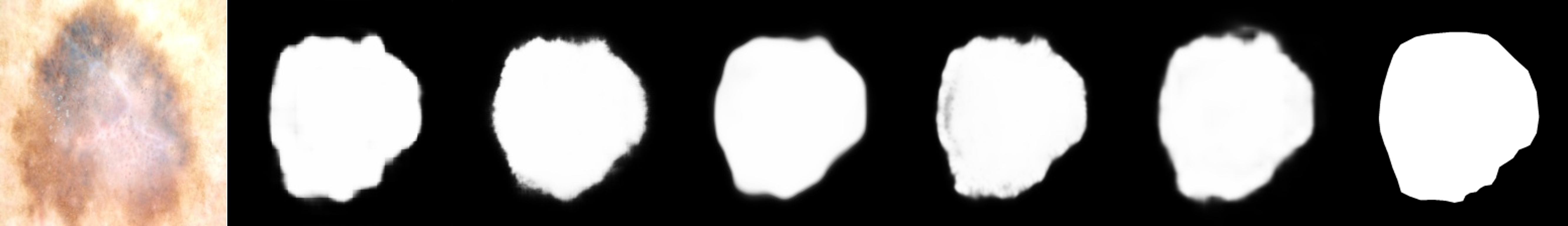}}
  \centerline{(C) skin}\medskip
\end{minipage}
\caption{Qualitative comparison of different segmentation results.}
\label{fig:res}
\end{figure*}

\textbf{Segmentation results.} In this section, we presented qualitative results on three data sets and compared with other SOTA semantic segmentation networks to prove the superior performance of our NPNet. We set up the same parameters for the same data set in different network models, and all models were trained from scratch. Since U-net is still the baseline of many networks, we also introduce several SOTA models based on U-net for comparison. Table~\ref{table1} shows that our model is superior to other SOTA models in terms of performance and is 1/50 of U-net in terms of model size and computation costs. In all the figures demonstrating the qualitative results in Fig~\ref{fig:res}, the sequence are origin image, FCN8s, SegNet, PSPNet, UNet, NPNet, mask, respectively.
It can be found from these figures that the model proposed can effectively reduce the loss of information with our non-pooling module, so as to retain more details and achieve better performance. Moreover, the model size and computation cost of this paper are only 1/100 of U-net++.\\
\textbf{Ablation Studies.} The attention mechanism module in SENet plays an important role, and many models insert this template into their models to achieve better performance. Therefore, we conducted an experimental comparison on three data sets of our proposed attention enhancement module. The difference between the model size and the computation costs of these three models can be negligible, the experimental results in Table~\ref{table2} prove that our attention enhancement module is better than SENet as a plug-and-play attention module.

\begin{table}[htp]
    \caption{Evaluation of proposed attention enhancement module}
    \centering
    \begin{tabular}{cccccccc}
        \hline
        Dataset& Methods&Attention & IOU& Dice\\
        \hline
        CVC  & NPNet & no & 0.7439 & 0.8157\\
        & NPNet & senet & 0.7448 & 0.8186 \\
        & NPNet & cam & \textbf{0.7766} & \textbf{0.8397}\\
        \hline
        luna  & NPNet & no & 0.9758 & 0.9807\\
        & NPNet & senet & 0.9772 & 0.9820\\
        & NPNet & cam & \textbf{0.9785} & \textbf{0.9832}\\
        \hline
        skin  & NPNet & no & 0.8131 & 0.8742\\
        & NPNet & senet & 0.8091 & 0.8709\\
        & NPNet & cam & \textbf{0.8165} & \textbf{0.8766}\\
        \hline       
    \end{tabular}
    \label{table1}
\end{table}

\section{Conclusion}\label{sec5}
In this work, we propose a novel semantic segmentation network with non-pooling operation for the first time, which can effectively alleviate the problem of information loss and difficult recovery caused by the pooling operation. Our proposed network also gets rid of common encoding and decoding structures. In addition, we also proposed an attention module to enhance feature information, which can be easily inserted into other network models with fewer parameters, Experiment results on three datasets show that our model can surpass state-of-art counterparts with lightweight parameters and MACs.

%
%


%
%

\bibliographystyle{spbasic}      
\bibliography{sn-bibliography}   


\end{document}